\documentclass[letterpaper, 10 pt, conference]{ieeeconf}  

\IEEEoverridecommandlockouts                              

\usepackage{graphicx} 
\usepackage{epsfig} 
\usepackage{mathptmx} 
\usepackage{times} 
\usepackage{amsmath} 
\usepackage{amssymb}  
\usepackage{subfigure}
\title{\LARGE \bf Anti-Jerk On-Ramp Merging Using Deep Reinforcement Learning}

\author{Yuan Lin$^{1}$, John McPhee$^{2}$, and Nasser L. Azad$^{3}$
\thanks{$^{1}$Dr. Yuan Lin is a Postdoctoral Fellow in the Systems Design Engineering Department at the University of Waterloo, Ontario, Canada N2L 3G1.
        {\tt\small y428lin@uwaterloo.ca}}%
\thanks{$^{2}$Dr. John McPhee is a Professor and Canada Research Chair in the Systems Design Engineering Department at the University of Waterloo, Ontario, Canada N2L 3G1.
        {\tt\small mcphee@uwaterloo.ca}}%
\thanks{$^{3}$Dr. Nasser L. Azad is an Associate Professor in the Systems Design Engineering Department at the University of Waterloo, Ontario, Canada N2L 3G1.
        {\tt\small nlashgarianazad@uwaterloo.ca}}%
}

\begin{document}

\maketitle
\thispagestyle{empty}
\pagestyle{empty}

\begin{abstract}

Deep Reinforcement Learning (DRL) is used here for decentralized decision-making and longitudinal control for high-speed on-ramp merging. The DRL environment state includes the states of five vehicles: the merging vehicle, along with two preceding and two following vehicles when the merging vehicle is or is projected on the main road. The control action is the acceleration of the merging vehicle. Deep Deterministic Policy Gradient (DDPG) is the DRL algorithm for training to output continuous control actions. We investigated the relationship between collision avoidance for safety and jerk minimization for passenger comfort in the multi-objective reward function by obtaining the Pareto front. We found that, with a small jerk penalty in the multi-objective reward function, the vehicle jerk could be reduced by 73\% compared with no jerk penalty while the collision rate was maintained at zero. Regardless of the jerk penalty, the merging vehicle exhibited decision-making strategies such as merging ahead or behind a main-road vehicle.

\end{abstract}

\section{INTRODUCTION}

Automated vehicle development is important work that could improve transportation safety and mobility. There are commercially available Advanced Driver Assistance Systems such as Adaptive Cruise Control, Lane Keeping Assistance, Blind Spot Warning, and Driver Drowsiness Detection \cite{eskandarian2012}. Some companies are working on highly automated vehicles that can perform complex automated driving tasks such as merging, intersection traversing, and roundabout maneuvers. These intelligent driving functions assist human drivers but safety is yet to be guaranteed. Vehicle-to-everything (V2X) wireless communications could enhance safety and mobility \cite{blum2004}, but 100\% penetration of V2X communications for every vehicle and road network in the world is still in the future.

Merging is a challenging task for both human drivers and automated vehicles. According to the US Department of Transportation, nearly 300,000 merging accidents happen every year with 50,000 being fatal \cite{merging_accidents}. The pilot vehicles of the leading self-driving car company, Waymo, were reportedly unable to merge autonomously \cite{waymo}. Merging scenarios have many variations: (1) There are low-speed merging in urban driving and high-speed freeway on-ramp merging. In urban driving, the automated vehicle may stop before the main road while in freeway merging, stopping is dangerous. Intuitively, high-speed merging seems to be riskier since there is less time for reaction given the limits of vehicle dynamics. (2) For freeway on-ramps, there are parallel-type ramps wherein a part of the ramp is connected and parallel to the main road such that the merging vehicle could switch to the main road at any point of the parallel portion of the ramp. There are also taper-type ramps wherein the ramp is only connected to the main road at the ramp's end. Parallel-type ramps are preferred by the Federal Highway Administration. However, there is still significant presence of taper-type ramps in the world \cite{aashto2001policy}. (3) The main road traffic may be light or dense. In extremely dense traffic, the interactions with manually driven vehicles may be crucial for successful merging \cite{sun2019interpretable}. (4) There are also centralized and decentralized merging. Centralized merging relies on roadside centralized controllers to coordinate the vehicles using V2X communications, while in decentralized merging, the merging vehicle merges completely on its own. In this work, we consider decentralized high-speed non-stop merging into freeways with moderate traffic via taper-type ramps.

In the current literature, there are rule-based and optimization-based methods proposed to tackle the automated merging problem \cite{rios2016survey}. The rule-based approaches include heuristics. The optimization-based approaches require modeling merging in a multi-agent system framework. In \cite{rios2017automated}, the authors used an optimal control approach for centralized merging control with V2X communications. In \cite{cao2015cooperative}, the authors used Model Predictive Control (MPC) to control an automated vehicle to merge via a parallel-type on-ramp in a decentralized fashion. In \cite{wang2018agent}, the authors used the virtual platoon method through centralized control to allow vehicles to merge with a predetermined sequence. Optimization-based methods require online optimization which could become an obstacle for realtime application if the number of vehicles involved is large \cite{maitland2020quasi,ding2019rule}. In \cite{kang2017game}, the authors investigated merging decision-making using game theory.

In recent years, there are numerous studies that investigate automated merging using DRL, which is a learning-based method. This is because DRL has demonstrated super-human performance in playing complex board games \cite{silver2016mastering}. Additionally, DRL has been shown to achieve near-optimal control performance when compared to MPC \cite{lin2019comparison}. DRL utilizes deep (multi-layer) neural nets as the policy approximators, which require very little computation time during deployment. In \cite{wang2018autonomous}, the authors used reinforcement learning to train a continuous action controller for high-speed merging with a quadratic function with neural-net weights as the policy approximator; the objectives include maintaining a safe distance with neighboring vehicles while minimizing control effort; however, the merging decision-making such as gap selection and the merging trajectories are not shown in the paper. In \cite{nishi2019merging}, the authors used the passive actor-critic algorithm for gap-selection and continuous action control for merging into dense traffic; the rewards encourage merging midway between vehicles on the main road while keeping the same speed as the preceding vehicle. In \cite{bouton2019cooperation}, the authors considered the cooperation levels of surrounding vehicles and used Deep Q-Networks for discrete action control of the merging vehicle; urban driving with dense traffic was considered and the vehicle speeds were around 5m/s. In \cite{hu2019interaction}, the authors also considered the cooperation levels of vehicles and developed a multi-agent reinforcement learning algorithm for discrete action control; vehicle speeds of 15m/s and 20m/s were considered and the algorithm achieved zero collision rate in the simulation; the authors also observed the trained decision-making strategies of merging ahead and behind a main-road vehicle.

Automotive vehicle control usually involves multiple objectives, which may include safety, passenger comfort, and energy efficiency \cite{wei2004objective}. These objectives may be presented as penalizing collision, jerk, and control action, respectively, in the multi-objective reward or cost function. Safety is the priority concern while the other objectives are also important. For example, passenger comfort is related to passengers' well-being and thus impacts the public acceptance of automated vehicles. However, a large number of published papers on automated vehicle control using DRL focus on safety and neglect the other objectives. For examples, vehicle jerk and control action penalties are not included in papers that use DRL for merging \cite{bouton2019cooperation,nishi2019merging}, intersection \cite{qiao2018automatically,qiao2018pomdp,isele2018navigating,li2019urban}, and roundabout \cite{jang2019simulation} maneuvers. There is also limited work that considers anti-jerk control in the DRL objectives. In \cite{hu2019interaction}, the authors exclude actions that lead to high jerk values during training to obtain a merging policy. To our best knowledge, there is no work that systematically investigates the impact of jerk minimization on the safety compromise of automated vehicle control using reinforcement learning. The systematic study of the interaction of multiple objectives involves obtaining the Pareto front, wherein no solution can improve at least one objective without degrading any other objective \cite{ngatchou2005pareto}. A typical way to obtain the Pareto front is to vary the ratios of the weights for different objectives \cite{van2014multi}.

Our work here studies decentralized decision-making and longitudinal control for merging using DRL. The decision-making is defined as merging gap selection, which results from merging ahead or behind. The decision-making of merging ahead or behind is not a direct output from DRL; instead, it is an outcome of controlling the acceleration of the merging vehicle via DRL. The main contributions of this work are two-fold. Firstly, we use DDPG to train the merging policy. DDPG is based on a deterministic policy and outputs continuous actions for decision-making and control \cite{lillicrap2015continuous}. There is no published paper that utilizes DDPG for merging. Secondly, we obtain the Pareto front for the objectives of collision and jerk minimizations by varying the weight of the jerk penalty. It allows us to investigate if jerk could be reduced while maintaining zero collisions. It also enables us to study the trajectory smoothness and decision-making strategies without and with jerk minimization.

The rest of the paper is organized as follows: in Section II, the preliminaries of reinforcement learning and DDPG are introduced; in Section III, the merging problem is formulated and cast into the reinforcement learning framework; in Section IV, training and testing simulations are presented and results are evaluated; in Section V, we draw conclusions and present possible future work.

\section{REINFORCEMENT LEARNING PRELIMINARIES}

In this section, we introduce the preliminaries of reinforcement learning and DDPG.

\subsection{Reinforcement Learning}

Reinforcement learning is a learning-based method for decision-making and control. In reinforcement learning, an agent takes an action based on the environment state at the current time step, and the environment subsequently moves to another state at the next time step. The agent also receives a reward based on the action taken. The action and reward are based on probabilities. Reinforcement learning algorithms seek to minimize the expected discounted cumulative reward for each episode. Specifically, the discounted cumulative reward for a state-action pair is called the Q-value, i.e. $Q(s_t,a_t) = E[\sum_{\tau=t}^{\tau=T} \gamma^{\,\tau-t} r(s_{\tau},a_{\tau})]$ where $r(s_{\tau},a_{\tau})$ is the reward for the state $s$ and action $a$ at time step $\tau$, and $\gamma \in [0,1]$ is the discount factor. The reinforcement learning problem is solved via Bellman's principle of optimality which means that, if the optimal Q-value for the next step is known, then action for the current time step must be optimal. That is, $Q^\star(s_{t},a_{t}) = r(s_t,a_t) + \gamma Q^\star(s_{t+1},a_{t+1})$ for an optimal policy, with $^\star$ denoting the optimality.

\subsection{Deep Deterministic Policy Gradient}

There are different DRL algorithms available, while we use DDPG for continuous control. There are two networks in DDPG: actor and critic networks. The critic network represents the Q-value which is the discounted cumulative reward. The critic network is iteratively updated based on Bellman's principle of optimality by minimizing the root-mean-squared loss $L_t = r(s_t,a_t) + (1-I)\gamma Q(s_{t+1},\mu(s_{t+1}|\theta^\pi)) - Q(s_t,a_t|\theta^Q)$ using gradient descent where $\theta^Q$ denotes the critic neural net weights, and $\theta^\pi$ denotes the actor neural net weights. The $I\in$\{0,1\} is the indicator for episode termination with $I=1$ means termination and $I=0$ means that the episode is not yet terminated. The actor network is the policy network that maps the environment state to action. The actor network is learned by performing a gradient ascent on the Q-value $Q(s_t,\mu(s_t|\theta^\pi))$ with respect to actor network parameters $\theta^\pi$.

Several techniques are necessary to facilitate training and improve training stability. Those include target networks, mini-batch gradient descent, and experience replay. During training, Gaussian noise is added to the action for exploration purpose. Table~\ref{table:ddpg_para} shows the DDPG parameter values that are used for the merging problem described next. These DDPG parameters are tuned through trial and error. Both the actor and critic networks have 2 hidden layers with 64 neurons for each layer.

\begin{table}[h]
\caption{DDPG parameter values.}
\begin{center}
\begin{tabular}{|c|c|}
\hline
Target network update coefficient & 0.001\\
\hline
Reward discount factor $\gamma$ & 0.99\\
\hline
Actor learning rate & 0.0001\\
\hline
Critic learning rate & 0.001\\
\hline
Experience replay memory size & 1500000\\
\hline
Mini-batch size & 128\\
\hline
Actor Gaussian noise mean & 0\\
\hline
Actor Gaussian noise standard deviation & 0.02\\
\hline
\end{tabular}
\end{center}
\label{table:ddpg_para}
\end{table}

\section{MERGING PROBLEM FORMULATION}

\subsection{Merging Environment}

The merging environment is created in the Simulation of Urban Mobility (SUMO) driving simulator \cite{lopez2018microscopic}. We consider a vehicle seeking to merge onto a freeway (main road) via a taper-type on-ramp. The merging vehicle is an automated vehicle equipped with a suite of perception sensors such as lidar and radar. We consider the horizontal sensing range of the merging vehicle as a circle with a radius of 200 meters, which could be made possible by advanced sensors \cite{fersch2017challenges,hecht2018lidar}. The automated vehicle is assumed to perceive the states of all the vehicles (including itself) within its sensing range and perform decentralized merging control. There is no centralized control for the merging and main-road vehicles via wireless communications. The merging vehicle is considered as a point mass and utilizes a kinematic model for state update. That is, there is no vehicle dynamics or delay considered for the merging vehicle.

\begin{figure}[htbp]
\centering
\includegraphics[width=3in]{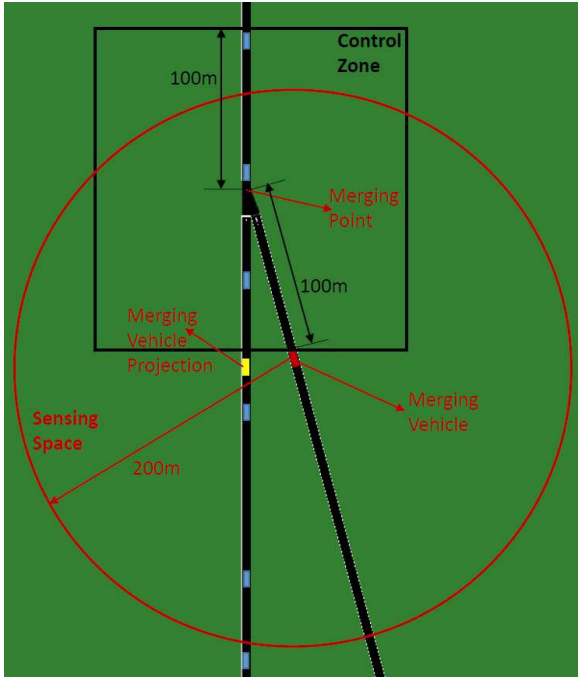}
\caption{Schematic for merging.}
\label{fig:schematic}
\end{figure}

The main-road vehicles can perform car following and collision avoidance based on the Intelligent Driver Model (IDM) \cite{treiber2000congested}. They can slow down when the merging vehicle enters the small junction area that connects the on-ramp and the main road, see Fig.~\ref{fig:schematic}. Main-road vehicles have speeds that are near the speed limit $v_{limit} = 29.06m/s$ (65mph). Particularly, each main-road vehicle has a desired speed as $v_{limit}*\gamma$ where $\gamma$ is constant for a vehicle. The $\gamma$ is a Gaussian distribution of mean 1 and standard deviation 0.1 with values clipped within [0.8, 1.2]. We consider moderate traffic density such that a main-road vehicle is generated at the far bottom of the main road with a probability of 0.5 at every second. Due to the probabilistic traffic generation, the speed variation among vehicles, and IDM model parameter variations (such as car-following headway difference), the main-road vehicles have very different inter-vehicular distance gaps near the merging junction. All the main-road vehicles have normal acceleration values in [-4.5, 2.6]m/s$^2$. When emergency situations happen, such as the merging vehicle merges too closely, the following vehicle can decelerate further to the minimum -9m/s$^2$ which is the emergency braking deceleration. These definitions for the main-road vehicles remain the same for DRL training and testing.

We define a control zone for the merging vehicle that is 100m behind the merging point on the on-ramp and 100m ahead of the merging point on the main road, see Fig.~\ref{fig:schematic}. The DRL policy is for controlling the merging vehicles only in the control zone. We assume that effective decisions could be made within 100m behind the merging point since the merging vehicle could come to a complete stop within 100m given the speed limit in case that merging is not successful. Additionally, the beginning portion of an on-ramp is designed for the merging vehicle to accelerate from low speeds and merging decisions may not be made during the initial acceleration phase in practice. The 100m ahead of the merging point is used to evaluate the merging success since collision could occur after merging if the merging vehicle does not merge with appropriate speeds and gaps between vehicles.

We assume that there is no other on-ramp merging vehicle in front of the controlled merging vehicle within the 100m behind the merging point. The main road is assumed to be single-lane such that there is no lane change behavior by the main-road vehicles. Each simulation time step is 0.1 seconds.

For each merging episode, the merging vehicle is initially positioned at 100m behind the merging point on the on-ramp. The initial velocity of the merging vehicle is randomly distributed in [22.35, 26.82]m/s (50-60mph). This is based on the suggestion of the US Department of Transportation that a merging vehicle reaches at least 50mph before merging to a freeway with 65mph speed limit \cite{aashto2001policy}. The initial acceleration of the merging vehicle is zero. When stops, collisions, or merging successes happen, the corresponding episodes are terminated; the merging vehicle is then deleted and regenerated with new initial conditions. Merging success is defined as merging past the control zone with no stop or collision regardless of braking by the main-road vehicles.

Based on the merging environment, we design the following reinforcement learning framework for training a merging policy.
\subsection{Environment State}
The environment state of the reinforcement learning framework includes the states of five vehicles: the merging vehicle ($m$), along with its two preceding ($p1$, $p2$) and two following ($f1$, $f2$) vehicles when the merging vehicle is or is projected on the main road, see Fig.~\ref{fig:schematic}. The merging vehicle's projection on the main road has the same distance to the merging point as the merging vehicle on ramp. The state of the merging vehicle includes the distance to the merging point $d_m$, velocity $v_m$, and acceleration $a_m$. The states of the preceding and following vehicles include the distances to the merging point ($d_{p2}, d_{p1}, d_{f1}, d_{f2}$) and velocities ($v_{p2}, v_{p1}, v_{f1}, v_{f2}$). The acceleration values of the preceding and following vehicles are not included in the environment state since the acceleration measurements are usually noisy, especially when the vehicles are further away. The distance to the merging point is positive and negative before and after the merging point, respectively. The distance to the merging point is measured from the front bumper of the vehicle. The environment state is denoted as
\begin{equation}
\begin{split}
s = [d_{p2}, v_{p2}, d_{p1}, v_{p1}, d_m, v_m, a_m, d_{f1}, v_{f1}, d_{f2}, v_{f2}]
\end{split}
\label{eq:states}
\end{equation}
When there are fewer than two preceding or two following vehicles detected, virtual vehicles are assumed at the intersections of the sensing space and the main road to deliberately construct the five-vehicle state vector. This is a conservative approach for safety as there could be vehicles just outside the sensing space in real-world driving. The virtual vehicles have velocity values as the main road speed limit. Note that a vehicle is determined to be preceding or following relative to the merging vehicle or its projection. A following vehicle may become a preceding vehicle if the merging vehicle slows down on ramp and its projection moves behind the main-road vehicle.

\subsection{Control Action}
The control action of the reinforcement learning framework is the acceleration to the merging vehicle $a_m$. As we don't consider vehicle dynamics and delays, the acceleration control input is executed instantaneously by the merging vehicle. The velocity and position updates of the merging vehicle follow an Euler forward discretization with 0.1s as the time step. The acceleration control input for the merging vehicle is within [$a_{min}$,$a_{max}$] = [-4.5,2.6]m/s$^2$, which is the same normal acceleration range of a main-road vehicle. There is no emergency braking acceleration defined for the merging vehicle.

\subsection{Reward}
The reward at each time step includes:

1) After the merging point and until the end of the control zone (100m ahead of the merging point), the merging vehicle should be midway between the first preceding and first following vehicles, with the average speed of the two vehicles. The corresponding penalizing reward is defined as
\begin{equation}
\begin{split}
r_m = -w_m*(|w| + \frac{|(v_{p1}+v_{f1})/2 - v_m|}{\Delta v_{max}})
\end{split}
\label{eq:r_m}
\end{equation}
where $w_m$ is the weight for merging midway and $w$ is the midway ratio defined as the ratio between the difference and sum of the distance gaps among the merging, its first preceding, and its first following vehicles. The ratio $w$ has values in [0,1] wherein 0 means the distance gap between the merging and the first preceding vehicles is the same as the one between the merging and the first following vehicles, and 1 means the merging vehicle has zero distance gap with either the first preceding or first following vehicle. The $\Delta v_{max}$=5m/s is the maximum allowed speed difference that we defined.
\begin{equation}
\begin{split}
w = \frac {|d_{p1} - d_m - l_{p1}| - |d_m - d_{f1} - l_m|} {|d_{p1} - d_{fl} - l_{p1} - l_m|}
\end{split}
\label{eq:midway_ratio}
\end{equation}
where $l_{p1}$=5m, $l_m$=5m, and $l_{f1}$=5m are the vehicle lengths of the first preceding, merging, and first following vehicles, respectively.

2) When the first following vehicle performs braking $a_{f1}<0$m/s$^2$ in the control zone, a penalizing reward is given as:
\begin{equation}
\begin{split}
r_b = -w_b*\frac{|a_{f1}|} {max(|a_{min}|,a_{max})}
\end{split}
\label{eq:r_b}
\end{equation}
where $w_b$ is the weight for penalizing braking by the first following vehicle.

3) To reduce the jerk of the merging vehicle for passenger comfort, we define a penalizing reward as:
\begin{equation}
\begin{split}
r_j = -w_j*\frac {|j_m|} {j_{max}} = -w_j*\frac {|\dot a_m|} {j_{max}}
\end{split}
\label{eq:r_b}
\end{equation}
where $w_j$ is the weight for the jerk penalty. The $j_{max}$=3m/s$^3$ is the maximum allowed jerk value for passenger comfort \cite{batra2019real}. The $w_j$ value is varied in this work to obtain the Pareto front for the collision and jerk minimization objectives.

4) When the merging vehicle stops, a penalizing reward of -0.5 is given and the episode is terminated.

5) When the merging vehicle collides with any vehicle, a penalizing reward of -1 is given and the episode is terminated. A collision is registered when the inter-vehicular distance gap is less than 2.5m. Note that a larger penalty is given for collisions than for stops because collisions are considered more catastrophic.

6) When the merging vehicle successfully passes the control zone (100m ahead of the merging point), a reward of 1 is given.

We don't penalize the time it takes the agent to finish an episode since time minimization is inherent in the reinforcement learning discount factor $\gamma=0.99$ \cite{bouton2019cooperation}. In addition, penalizing stops during merging also contributes to time minimization. Table~\ref{table:para} shows the parameter values of the merging vehicle.

\begin{table}[h]
\caption{Parameter values for the merging vehicle.}
\begin{center}
\begin{tabular}{|p{5.5cm}|c|}
\hline
Minimum acceleration $a_{min}$ & -4.5m/s$^2$\\
\hline
Maximum acceleration $a_{max}$ & 2.6m/s$^2$\\
\hline
Weight for merging midway $w_m$ & 0.015\\
\hline
Maximum allowed speed difference $\Delta v_{max}$ & 5m/s\\
\hline
Weight for penalizing braking by first following vehicle $w_b$ & 0.015\\
\hline
Weight for penalizing jerk $w_j$ & 0 to 0.015\\
\hline
Maximum allowed jerk value $j_{max}$ & 3m/s$^3$\\
\hline
\end{tabular}
\end{center}
\label{table:para}
\end{table}


\begin{figure}[htbp]
\centering
\includegraphics[width=3.4in]{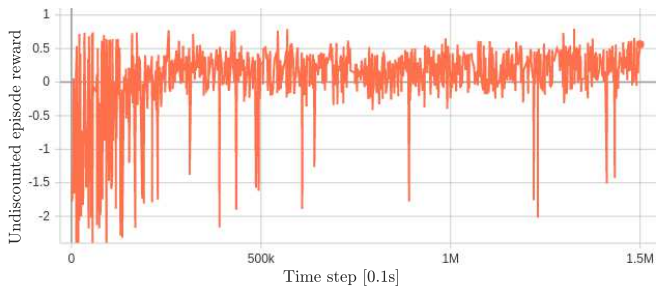}
\caption{Undiscounted episode reward during training.}
\label{fig:reward}
\end{figure}

\section{RESULTS}

In this section, we show and evaluate the training and testing simulation results for each weight for jerk penalty. Additionally, representative episodes are presented to show the merging vehicle's learned behaviors.

\subsection{Training}

For each weight for jerk penalty, we train the merging vehicle for 1.5 million simulation time steps, wherein we observe reasonable convergence of the undiscounted episode reward. Note that, before each episode starts, there is a 10-second buffer for initialization of the main-road traffic. Fig.~\ref{fig:reward} shows the undiscounted episode reward when the weight for jerk penalty $w_j$=0.015, which is the largest among the weights considered; the training convergence results for other weights look similar and are not plotted here. In general, at the initial phase of training, the DRL agent had episode rewards less than -0.5 very often, indicating many stops or collisions. As training progresses, the episode rewards less than -0.5 happened less often. It took around 10 hours for either training or testing on a computer with a 16-core (32-thread) AMD processor and a Nvidia GeForce RTX GPU.

\subsection{Testing}

For each weight for jerk penalty, the trained policy is tested for another 1.5 million simulation time steps which represent roughly 19000 merging episodes. We define the following metrics to evaluate the merging vehicle's performance during testing.

(1) Average collision rate: the number of collisions divided by the number of episodes.

(2) Average jerk: the mean of the average jerk magnitude of each episode.

(3) Average acceleration: the mean of the average acceleration magnitude of each episode.

(4) Average velocity: the mean of the average velocity of each episode.

(5) Merge-behind rate: the number of times of merging behind divided by the number of episodes.

(6) Merge-ahead rate: the number of times of merging ahead divided by the number of episodes.

Note that merging ahead or behind is defined as merging ahead or behind the first following vehicle.

\begin{figure*}[htbp]
\centering
\includegraphics[width=7in]{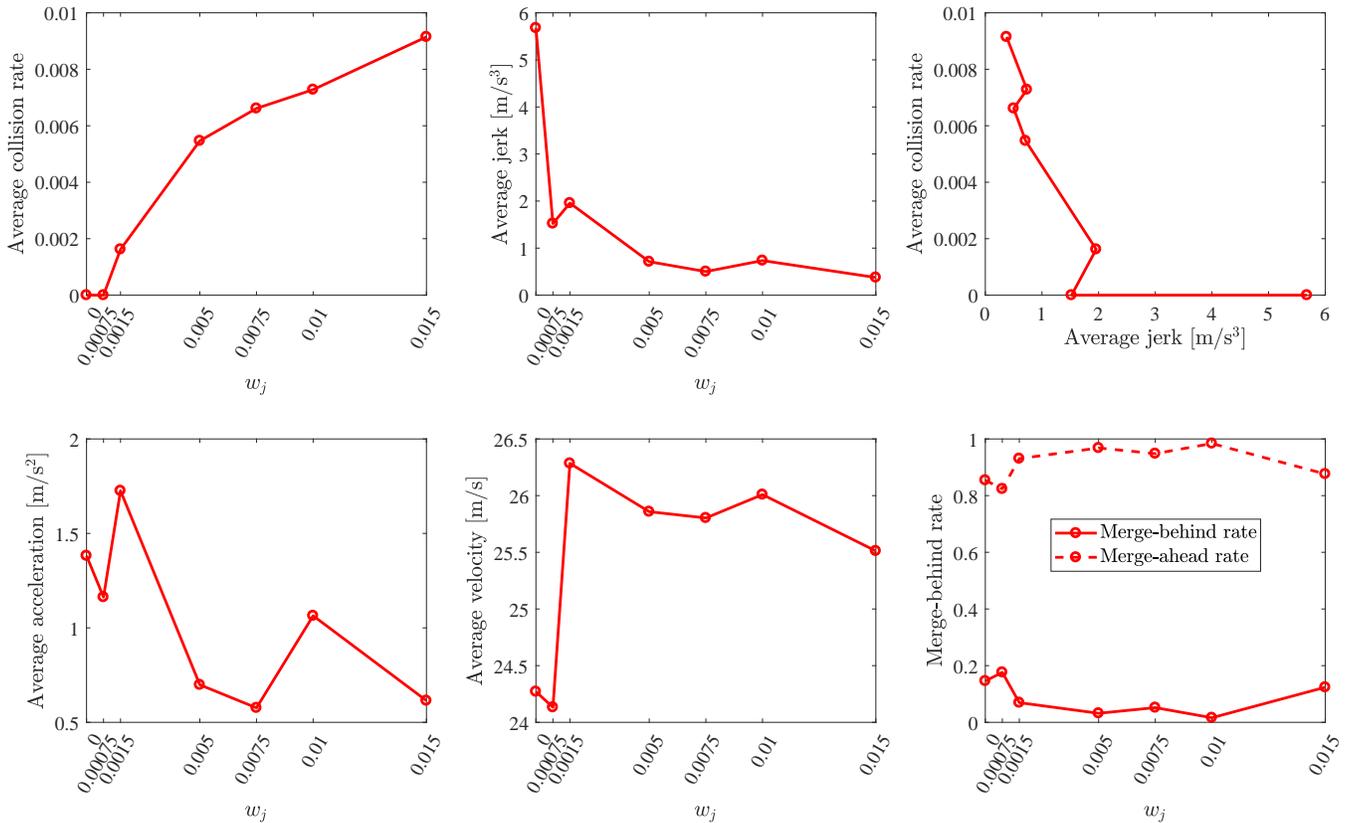}
\caption{Merging performance for different weights for jerk penalty $w_j$ during testing.}
\label{fig:pareto}
\end{figure*}

Fig.~\ref{fig:pareto} shows a summary of the metrics results for different weights for the jerk penalty. As DDPG is not robust and may converge to different sub-optimal policies each time, the plots in Fig.~\ref{fig:pareto} do not show smooth monotonic trends.

The plot of the average collision rate shows that when $w_j\leq$0.00075, there are zero collisions. When $w_j>$0.00075, the average collision rate increases with the increasing weight for jerk penalty. In contrast, the average jerk shows a decreasing trend with the increasing weight for jerk penalty. In particular, the average jerk is 1.52m/s$^3$ when $w_j$=0.00075 and 5.68m/s$^3$ when $w_j$=0, which indicates 73\% jerk reduction even though both cases have zero collisions. The Pareto front for the objectives of collision avoidance and jerk minimization is shown in the third plot of the first row of Fig.~\ref{fig:pareto}.

The average acceleration seems to decrease with the weight for jerk penalty although the trend is not obvious. However, a decreasing acceleration trend is usually expected as we penalize jerk, which indicates better energy efficiency \cite{wei2004objective,batra2019real}. The average velocity for smaller $w_j$ seems to be smaller than for larger $w_j$. The merge-behind and merge-ahead rates do not change significantly with the increasing weight for jerk penalty. Additionally, the merge-ahead rate is much larger than the merge-behind rate, indicating that the merging vehicle preferred to maintain the initial positioning sequence during merging.

Note that there were a few stops for different weights of jerk penalty during testing. For each case of jerk penalty, there were also a few occasions wherein the merging vehicle merged behind or ahead twice in one episode.

\subsection{Representative Episodes}

\begin{figure*}[htbp]
\centering
\includegraphics[width=7in]{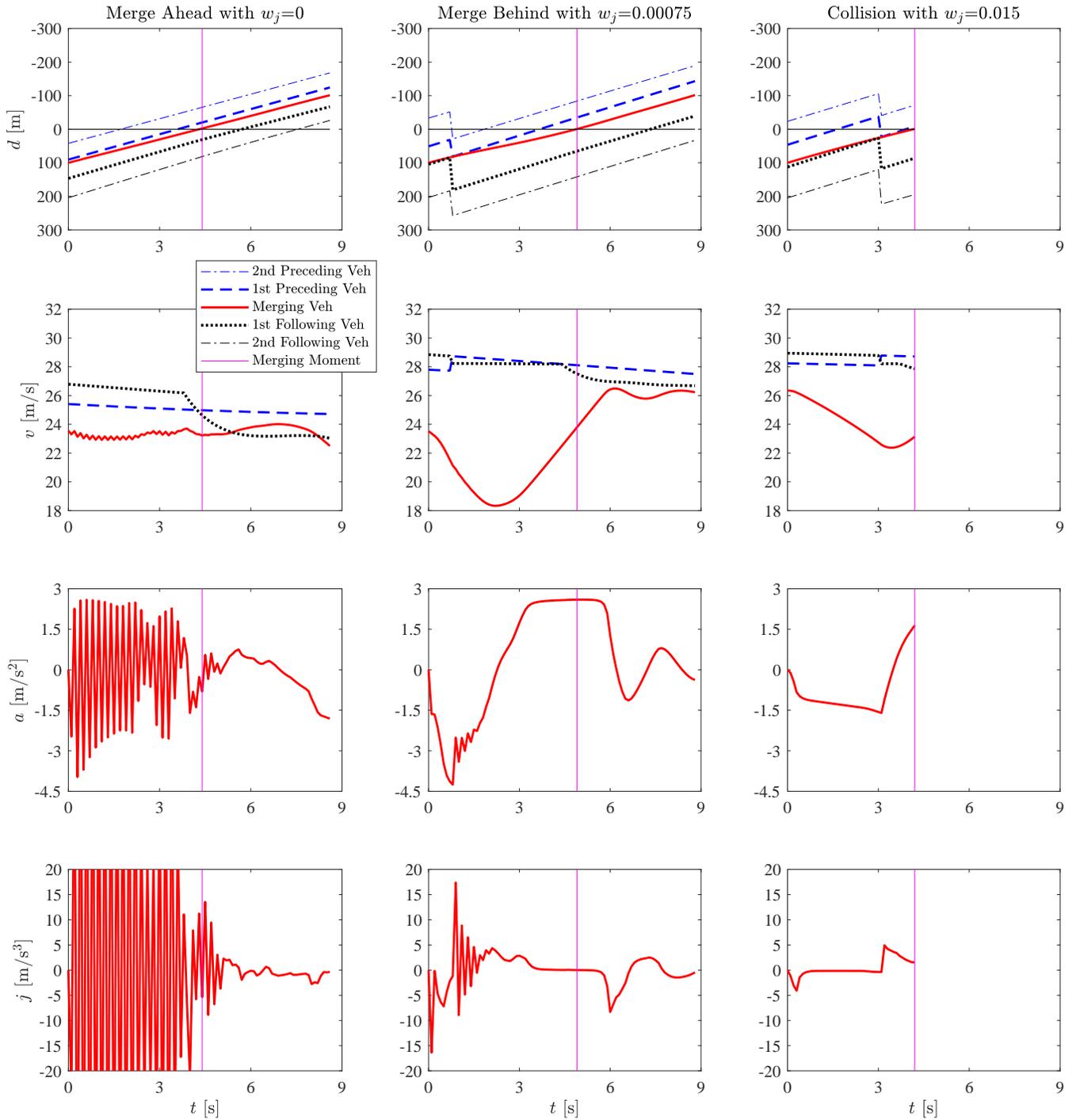}
\caption{Three representative episodes of merging during testing. Episode 1 (first column): merge ahead the first following vehicle with $w_j$=0; Episode 2 (second column): merge behind the first following vehicle with $w_j$=0.00075; and Episode 3 (third column): collision with $w_j$=0.015. In the second and third plots on the first row, the first following vehicle became the first preceding vehicle as the merging vehicle slowed down, which causes the sharp changes of the distance to merging point $d$ values for the related vehicles. In the first plot on the fourth row, the vehicle jerk values in the first 3 seconds are larger than 80m/s$^3$. While the lengths of the three episodes are different and not equal to 9s, the x-axis limit is set to 9s for comparison purposes.}
\label{fig:episodes}
\end{figure*}

To understand both the learned decision-making strategies and the impact of using different weights for jerk penalty on the trajectory smoothness, we plot the vehicle states under different decision-making strategies and for different weights for the jerk penalty, see Fig.~\ref{fig:episodes}. We plotted 3 episodes: Episode 1, the merging vehicle merged ahead with $w_j$=0; Episode 2, the merging vehicle merged behind with $w_j$=0.00075, and Episode 3, the merging vehicle caused a collision with $w_j$=0.015.

In Episode 1, the merging vehicle's initial projection on the main road was a little behind the first preceding vehicle and its speed was a little smaller than the first preceding vehicle. The merging vehicle's velocity was fairly constant during merging as the merging vehicle merged ahead the first following vehicle and remained behind the first preceding vehicle. The acceleration and jerk plots are very noisy because there is no penalty on the vehicle jerk.

In Episode 2, the merging vehicle's initial projection on the main road was a little ahead the first following vehicle and its speed was a little smaller than the first following vehicle. The trained policy enabled the merging vehicle to slow down significantly to merge behind the first following vehicle. The acceleration and jerk plots are much less jerky as compared with Episode 1, even with the small weight on jerk penalty $w_j$=0.00075. Note that the first following vehicle became the first preceding vehicle during merging due to the slowdown of the merging vehicle.

In Episode 3, the merging vehicle's initial projection on the main road was a little ahead the first following vehicle and its speed was a little smaller than the first following vehicle. The merging vehicle slowed down but not significantly. The first following vehicle became the first preceding vehicle before the merging vehicle collided with it in the front. This is likely due to the heavy weight for jerk penalty $w_j$=0.015, which leads to insufficient control action to avoid the collision.

\section{CONCLUSIONS}

In this work, we trained a freeway on-ramp merging policy using DDPG. The Pareto front for the merging objectives of collision avoidance and jerk reduction (passenger comfort) was obtained. We found that, with a relatively small jerk penalty with $w_j$=0.00075, the vehicle jerk could be significantly reduced by up to 73\% while merging could be maintained collision-free. As the weight for jerk penalty continues to increase, the gain of jerk reduction is not significant. Representative episodes show that when jerk penalty was not considered, the DRL-trained solution exhibited highly jerky jerk and acceleration profiles. The jerkiness may be in part due to the penalty spikes when stops or collisions happened which caused sharp neural-net weight updates during backpropagation. On the other hand, if the jerk penalty is too large, the merging vehicle could fail to execute sufficient control actions to avoid collisions.

At the decision-making level, the DRL-trained policies mainly exhibited two merging strategies: merging ahead or behind a following vehicle on the main road. Merging ahead dominated merging behind as it happened more than 80\% of the time. The resulting decision-making strategies are dependent on the rewards that we designed; merging ahead resulted in higher cumulative discounted rewards in the merging environment. We observed that merging ahead resulted in lower jerk magnitudes in general, which contributed to higher rewards. In addition, the time minimization due to the discount factor and the stop penalty may motivate merging ahead since it may take shorter time than merging behind with slowdown.

Even though our simulations showed zero collisions, guaranteed safety for the neural-net policy needs to be researched in the future work \cite{lee2019w}. Future work should also include considering vehicle dynamics in training since, in our previous work, we discovered that the kinematic-model-trained policy could cause significantly degraded performance in realistic situations with vehicle dynamics \cite{lin2019longitudinal}. Another research direction is to directly consider energy efficiency in the reward to train an economic merging policy \cite{vajedi2015ecological}.





\section*{ACKNOWLEDGMENT}

The authors thank Toyota, Ontario Centres of Excellence, and the Natural Sciences and Engineering Research Council of Canada for the support of this work. The authors appreciate the constructive feedback from Prof. Krzysztof Czarnecki and his research group at the University of Waterloo.

\bibliographystyle{IEEEtran}
\bibliography{references}

\end{document}